\documentclass[aps,11pt,tightenlines]{revtex4}
\usepackage{graphicx,amssymb,amsmath,bbm,amsfonts}

\newcommand{\1}{\leavevmode{\rm 1\ifmmode\mkern  -4.8mu\else\kern -.3em\fi I}}
\newcommand{\ket}[1]{\vert #1 \rangle}
\newcommand{\bra}[1]{\langle #1 \vert}

\newcommand{\braket}[2]{\langle #1 \vert #2 \rangle}

\newcommand{\ketbra}[2]{\vert #1 \rangle \! \langle #2 \vert}

\newcommand{\id}{{\mathbb I}}

\newcommand{\beq}{\begin{equation}}
\newcommand{\eeq}{\end{equation}}
\newcommand{\barr}{\begin{eqnarray}}
\newcommand{\earr}{\end{eqnarray}}
\newcommand{\andy}[1]{ }

\def\bra#1{\langle #1 |}
\def\ket#1{| #1 \rangle}

\newcommand{\Tr}{\mathop{\text{Tr}}\nolimits}

\begin{document}
\title{The discrimination problem for two ground states or two thermal
states of the quantum Ising model}
\author{Carmen Invernizzi}
\email{Carmen.Invernizzi@unimi.it} 
\affiliation{Dipartimento di Fisica, Universit\`a degli Studi di Milano, Italia}
\author{Matteo G. A. Paris}
\email{Matteo.Paris@unimi.it} 
\affiliation{Dipartimento di Fisica, Universit\`a degli Studi di Milano, Italia}
\affiliation{CNISM UdR Milano, I-20133 Milano, Italia}
\affiliation{ISI Foundation, I-10133 Torino, Italia}
\begin{abstract}
We address the one-dimensional quantum Ising model as an example of
system exhibiting criticality and study in some details the
discrimination problem for pairs of states corresponding to different
values of the coupling constant. We evaluate the error probability for
single-copy discrimination, the Chernoff bound for $n$-copy
discrimination in the asymptotic limit, and the Chernoff metric for the
discrimination of infinitesimally close states. We point out scaling
properties of the above quantities, and derive the external field
optimizing state discrimination for short chains as well as in the
thermodynamical limit, thus assessing criticality as a resource for
quantum discrimination in many-body systems. 
\end{abstract}
\maketitle
\section{Introduction}
In quantum state discrimination one should determine the state of a
quantum system based on the outcome of a certain measurement, and
assuming that the system may be prepared in a state chosen from a given
list of possible candidates.  Of course, when the candidate states are
not orthogonal, basic quantum mechanics dictates that no measurement can
distinguish perfectly between them. The objective is therefore to choose
some figure of merit characterizing the quality of the state
discrimination and optimize it over the space of allowed quantum
measurements.  This procedure, known as quantum state discrimination or
quantum hypothesis testing \cite{Helstrom,rv0,rv1,rv2}, plays a relevant
role in the characterization of signals and devices and, in turn, in the
development of quantum technology.
\par
The two main paradigms of state discrimination are unambiguous
identification \cite{ud1,ud2,ud3,ud4,ud5} and (ambiguous) minimum error
discrimination \cite{ad0,ad1,ad2,ad3,ad4,ad5}.  In the first method, to
discriminate among $N$ states one searches a measurement with $N+1$
outcomes, where the additional result accounts for inconclusive
detection, and in turn allows the conclusive determination in the
remaining cases.  On the other hand, in ambiguous discrimination one
looks for a measurement with $N$ outcomes, always leading to a
determination of the state, while accepting the possibility of a wrong
inference.  In this paper we restrict ourselves to the second method,
which basically consists in looking for the optimal measurement that
minimizes the probability of errors, {\em i.e} the overall probability
of a misidentification. For the discrimination between two states, pure
or mixed, the optimal measurement and the minimum error probability had
been derived by Helstrom \cite{Helstrom}. If $n$ copies of the system
are available the scaling of the error probability with the number
of copies may be expressed using the so-called quantum Chernoff bound 
$\xi_{QCB}$ \cite{PRLQCB,PRAQCB}. In particular, it has been proved that
$\xi_{QCB}$ defines a meaningful distinguishability measure when one has
to solve the problem of discriminating two sources that output many
identical copies of two quantum states. In addition, when considering
two states that are infinitesimally close, the quantum Chernoff bound
induces a metric on the manifold of quantum states. 
\par
In this paper we study the discrimination problem for two ground states 
or two thermal states of the Ising model in a transverse magnetic field, which 
represents a paradigmatic example of system which undergoes a second order 
quantum phase transition (QPT).  We consider the system both at zero 
and finite temperature, and address discrimination of states 
corresponding to different values of the coupling parameter.
In particular, we evaluate the error probability for single-copy 
discrimination, the Chernoff bound for $n$-copy discrimination in the 
asymptotic limit, and the Chernoff metric for the discrimination of 
infinitesimally close states.  We are 
interested in the scaling properties of the above quantities with the 
coupling itself, the temperature and the size of the system. Moreover, 
we look for the optimal value of the field that minimizes the probability 
of error and maximizes both the Chernoff bound and the corresponding metric. 
It turns out that criticality is a resource 
for quantum discrimination of states. Indeed, at zero temperature the 
critical point signs a minimum in the probability of error and a divergence 
in the QCB metric. Remarkably, despite the fact that Chernoff metric is 
associated to quantum discrimination and the Bures metric is related to 
quantum estimation \cite{pzmp,JIsing}, these different measures show the 
same critical behavior and carry the same information about the QPT of 
the system \cite{abasto}. 
\par     
The paper is organized as follows. In Section \ref{IsingModel}  we
introduce the model. In Section \ref{QSdiscrim} we review the basic
elements of quantum state discrimination and also illustrate the notion
of quantum Chernoff metric for the Ising model. In Section \ref{Tzero} we
study the distinguishability of states at zero temperature, both for the
case of few spins and then in the thermodynamic limit.     In Section
\ref{Tfinite} we consider the effects of temperature and the scaling
properties of the metric. Section \ref{conclusions} closes the paper with 
some concluding remarks.
\section{Quantum Ising model} \label{IsingModel}
We consider the one-dimensional Ising model of size $L$ as an example of
system which undergoes a zero-temperature quantum phase transition
\cite{SC1,SC2,SC3}.  The model is defined by the Hamiltonian
\begin{equation}\label{HIsing}
H= - J\sum_{k=1}^L \sigma_k^x\sigma_{k+1}^x - h\sum_{k=1}^L \sigma_k^z,
\end{equation}
where the $\sigma^\alpha_k$ are Pauli operators for the $k$-th site. We 
also assume periodic boundary conditions $\sigma_{L+1}^x = \sigma_{1}^x$. 
As the temperature and the field $h$ are varied one may identify different 
physical regions.  At zero temperature, the system undergoes a QPT for $h=J$ 
and becomes gapless. For $h<J$ the system is in an ordered phase whereas for 
$h>J$ the field  dominates, and the system is in a paramagnetic state. For 
temperature $T\ll \Delta$, $\Delta= \left| J-h\right|$ the system behaves 
quasi-classically,  whereas for $T\gg\Delta$ quantum effects dominate. The 
Hamiltonian (\ref{HIsing})  can be exactly diagonalized by a Bogoliubov 
transformation, leading to 
\begin{align}\label{eq:ising_diag}
H=\sum_{k>0} \Lambda_k \left ( \eta_k^\dagger\eta_k -1 \right),
\end{align}
where $\Lambda_k $ denotes the one particle energies and $\eta_k$ the
annihilation operator,
$\Lambda_{k}=\sqrt{\epsilon_{k}^{2}+\Delta_{k}^{2}}$, $\Delta_k = J\sin
(k)$, $\epsilon_k =(J\cos (k) +h)$.  
The one-particle excitations are created by the action of $\eta_k^\dagger 
= \cos (\frac{\theta_k}{2})d_k^\dagger + i \sin(\frac{\theta_k}{2})d_{-k} $ 
on the ground state
\begin{align}
\ket{\psi_0} = \bigotimes_{k}\left [ \cos\left (\frac{\theta_k}{2}\right)
\ket{00}_{k,-k} + i\sin \left (\frac{\theta_k}{2}\right )\ket{11}_{k,-k} \right ],
\end{align}
where $\vartheta_k = \tan^{-1} \frac{\epsilon_k}{\Delta_k}$ 
and $d_k\ket{00}_{k,-k}=d_{-k}\ket{00}_{-k,k}=\eta_k \ket{\psi_0}=0$.
Strictly speaking, Eq.~(\ref{eq:ising_diag}) holds in the sector with even
number of fermions. In this case, periodic boundary conditions on the
spins induce antiperiodic BC's on the fermions and the momenta satisfy
$k=\frac{(2n+1)\pi}{L}$. In the sector with odd number of particles,
instead, one has $k=\frac{(2n)\pi}{L}$ and one must carefully treat
excitations at $k=0$ and $k=\pi$.  In any case, the ground state of
(\ref{HIsing}) belongs to the even sector so that, at zero temperature
we can use Eq.~(\ref{eq:ising_diag}) for any finite $L$. At positive
temperature we will be primarily interested in large system sizes and
therefore we can neglect boundary terms in the Hamiltonian and use
Eq.~(\ref{eq:ising_diag}) in the whole Fock space. 
\section{Elements of quantum states discrimination}\label{QSdiscrim}
Suppose we have a quantum system which may be prepared in different states
$\rho_k$, $k=1,..,N$, chosen from a given set, with {\em a priori}
probability $z_k$, $\sum_k z_k =1$. A discrimination problem arises 
in any situation where the system is presented to an experimenter who 
has to infer the state of system by performing a measurement.
The states are known, as well as the {\em a priori} probabilities, 
but we don't know which state has been actually sent to the observer.
The simplest case occurs when the system may be prepared in two
possible states, described by the density matrices $\rho_1$ and
$\rho_2$, with {\em a priori} probabilities $z_1$ and $z_2=1-z_1$.
Any strategy for the (ambiguous) discrimination between these two states 
amounts to define a two-outcomes POVM $\{ E_1, E_2\}$ on the system,
where $E_1+E_2  =\id$ and $E_k \geq 0$ $\forall k$. After observing the 
outcome $j$ the observer infers that the state of the system is $\rho_j$. 
The probability of inferring the state $\rho_j$ when the true state is $\rho_k$ 
is thus given by $P_{jk} = \hbox{Tr}\left[\rho_k E_j\right]$ and the optimal 
POVM for the discrimination problem is the one minimizing the overall 
probability of a misidentification i.e. $P_e = z_1 P_{21} + z_2 P_{12}$. 
For the simplest case of equiprobable hypotheses ($z_1=z_2=1/2$) we have
$P_e = \frac{1}{2} \left(1-\Tr \left[E_2 \Gamma\right]\right)$ where $\Gamma=\rho_2
-\rho_1$. $P_e$ is minimized by 
choosing $E_2$ as the projector over the positive subspace of $\Gamma$. 
Then we have $\Tr [E_2\Gamma] =\Tr |\Gamma| $ and 
$P_e = \frac{1}{2}\left ( 1- \Tr \left|\Gamma \right|\right)$
where $|A|=\sqrt{A^\dag A}$. When $\rho_k=|\psi_k\rangle
\langle\psi_k|$ are pure states the error probability reduces to 
$P_e = \frac{1}{2}\left ( 1- \sqrt{1-|\langle\psi_1|\psi_2
\rangle|^2}|\right)$.
\par
Let us now suppose that $n$ copies of both $\rho_1$ and $\rho_2$ are 
available for the discrimination. The problem may be addressed
using the above formulas upon replacing $\rho$ with $\rho^{\otimes n}$. 
We thus need to analyze the  quantity $P_{e,n} = \frac{1}{2}\left ( 1
- \Tr |\rho_2^{\otimes n} -\rho_1^{\otimes n}|\right )$. It 
turns out that in the asymptotic limit of large $n$ the error probability 
decreases exponentially with $n$ as $P_{e,n} \sim e^{-n\xi_{QCB}}$
where the quantity $\xi_{QCB}$ is called the \emph{quantum Chernoff bound} 
(QCB) and may be evaluated as follows \cite{PRLQCB}
\begin{align}
\xi_{QCB} = -\log \min_{0\leq s\leq 1} \Tr\left[\rho_1 ^s\:
\rho_2^{1-s}\right]\:.
\end{align}
For pure states QCB achieves its superior limit, which is given 
in terms of the overlap between the two states $\xi_{QCB}= - 
\log |\langle\psi_1|\psi_2\rangle|^2$. The QCB 
introduces a measure of distinguishability for 
density operators which acquires an operational meaning in the
asymptotic limit. For a fixed probability of error $P_e$, the larger 
is the $\xi_{QCB}$, the smaller the number of copies of $\rho_1$ and 
$\rho_2$ we will need in order to distinguish them.
\par
Upon considering two nearby states $\rho$ and $\rho +d\rho$, the 
QCB induces the following distance over the manifold of quantum states 
\begin{align}
ds_{QCB} ^2:= 1 - \exp (-\xi_{QCB}) = \frac{1}{2}\sum_{m,n} 
\frac{|\bra{\varphi_m}d\rho\ket{\varphi_n}|^2}{(\sqrt{\rho_n}+\sqrt{\rho_m})^2}
\end{align}
where the $|\varphi_n\rangle$'s are the eigenvectors of $\rho= \sum_n \rho_n 
\ketbra{\varphi_n}{\varphi_n}$. In the following we will consider
infinitesimally close states obtained upon varying a Hamiltonian
parameter $\lambda$, and $d\rho$ will correspond to
$d\rho=\partial\rho/\partial\lambda\,d\lambda$. 
The above definition means that the bigger is the QCB distance, the
smaller is the asymptotic error probability of discriminating a 
given states from its close neighbors. 
\par
In the following we will consider discrimination for ground and thermal
states. In this case the eigenstates of $\rho$ are those of the Hamiltonian 
and the distance may be written as the sum of two contributions 
\begin{align}
ds_{QCB}^2 = & \underbrace{
\frac{1}{8}\sum_n  \frac{(d\rho_n)^2}{\rho_n}
}+ \underbrace{
\frac{1}{2}\sum_{n\neq m}\frac{|\braket{\varphi_n}{d\varphi_m}|^2
(\rho_n-\rho_m)}{(\sqrt{\rho_n}+\sqrt{\rho_m})^2}} 
\label{dsQCB} \\ 
& \quad\quad ds_c^2 \qquad\qquad\qquad\qquad\quad  ds_{nc}^2 \nonumber
\end{align}
where $ds_c^2$ refers to the classical part since it only depends on the 
Boltzmann weights of the eigenstates in the density operator, whereas $ds_{nc}^2$ 
to the nonclassical one because it explicitly depends on the dependence
of the eigenstates from the parameter of interest. If we consider the
Ising model of the previous Section and address discrimination of states
labeled by different values of the coupling $J$, the QCB distance can
be expressed by the metric $g_J$, $ds_{QCB}^2=g_J dJ^2$. We have \cite{abasto}
\begin{align}
 g_J =& 
\underbrace{ 
 \frac{\beta^{2}}{32}\sum_{k}\frac{\left(\partial_{J}
 \Lambda_{k}\right)^{2}}{\cosh^{2}\left(\beta\Lambda_{k}/2\right)} 
 }+\underbrace{
 \frac{1}{4}\sum_{k} \tanh ^2 (\beta \Lambda_k /2) 
 \left(\partial_{J}\vartheta_{k}\right)^{2}}
\label{dsB} \\ 
& \qquad\quad\quad g_J ^c  \qquad\qquad\qquad\qquad\qquad\quad
g_J^{nc}\nonumber 
\end{align}
Recent results about the Chernoff bound metric $ds_{QCB}^2$
\cite{abasto,scarola}
have shown that it may be used to investigate the phase diagram the Ising
model, i.e. to identify, in 
terms of different scaling with temperature, quasiclassical and
quantum-critical regions. 
These results extend recent ones obtained using the Bures metric 
$ds_B^2$ (or the fidelity) \cite{pzPRA07,ZH,HQZ}
{\em i.e}
\begin{equation}
ds_B^2 = \frac12 \sum_{nm} \frac{\left|\langle \varphi_m|d \rho 
| \varphi_n\rangle\right|^2}{\rho_n+ \rho_m}
\label{HH}\:.
\end{equation}
In turn, one has the relation 
$\frac12ds_B^2\leq ds_{QCB}^2 \leq ds_B^2$
which shows that the Bures and the QCB metric have the same divergent
behavior i.e. one metric diverges iff the other does.  Then one can
exploit the results on the scaling behavior of the Bures metric derived
in \cite{pzPRA07} to discriminate quantum states.  Moreover, in the
following we will see that when the system is in its ground state,
$ds_{QCB}^2 =ds_B^2$ whereas at increasing temperature $T$, $ds_{QCB}^2
\to \frac12 ds_B^2$. 
\section{Discrimination of ground states}\label{Tzero}
At zero temperature the system is in the ground state and the problem is
that of discriminating two pure states corresponding to two different
values $J_1$ and $J_2$ of the coupling $J$. The probability of error 
is given in terms of the overlap $|\braket{\psi_1}{\psi_2}|^2$, 
whereas the  minimum of $\Tr\left[\rho_1 ^s\:
\rho_2^{1-s}\right]$ reduces to the overlap itself since for 
pure states $\rho^s=\rho\; \forall s$. Thus the probability of error for 
the discrimination with $n$ copies scales as $P_{e,n} \sim
|\braket{\psi_1}{\psi_2}|^{2n}$. In other words, the QCB may be
expressed as $\xi_{QCB}=-\log\left[4\,P_e(1-P_e)\right]$.
In this section we address the discrimination problem at zero 
temperature by evaluating the probability of error and the 
QCB metric, pointing out scaling properties, and minimizing 
(maximizing) them as a function of the external field.
We first consider systems  made of few spins and then address 
the thermodynamic limit.
\subsection{Short Ising chains, $L=2, 3, 4$}
\begin{figure}[t]
\includegraphics[width=0.45\textwidth]{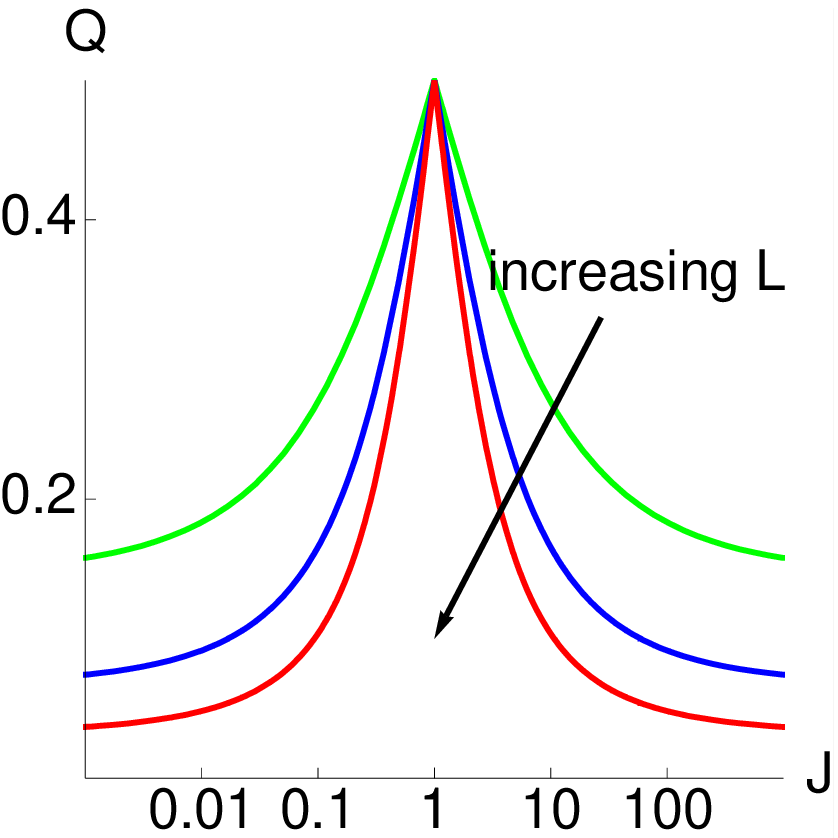}
\includegraphics[width=0.45\textwidth]{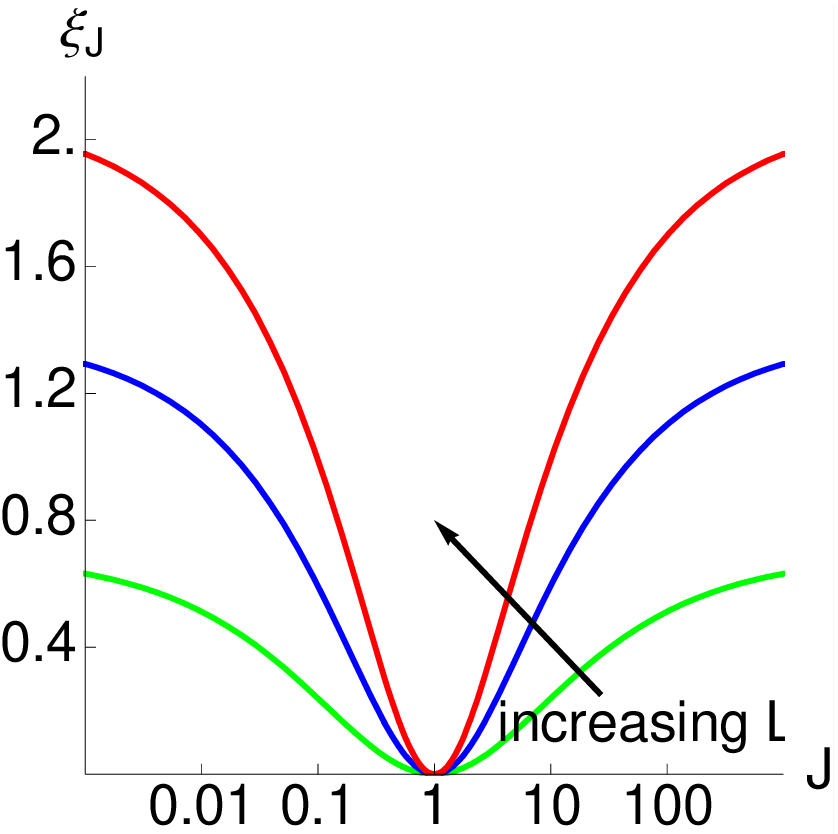}
\caption{(Left):Log-linear plot of the zero temperature rescaled minimum 
probability of error $Q(J)\equiv P_{e,min}(1,J,\sqrt{J})$ as a 
function of $J$ for $L=2,3,4$ (green, blue and red lines, respectively). 
The function has a cusp in $J=1$ and the two tails
go to zero faster with increasing size. According to the scaling in Eq. 
(\ref{sc1}) the relevant parameter is the ratio between the two 
couplings and not the absolute difference. In the log-linear plot, 
this means that $Q(J)$ is symmetric around $J=1$. (Right): The Chernoff
bound in the same conditions.}\label{Pe234}
\end{figure}

The probability of making a misidentification $P_e$ may be 
minimized by varying the value of the external field.
For the case 
$L=2,3,$ and $4$, $P_e$ is obtained by explicit diagonalization
of the Ising Hamiltonian. Minima of $P_e$ correspond to the field 
value $\tilde{h}=\sqrt{J_1 J_2}$, {\em i.e} the geometrical mean of 
the two (pseudo) critical values, and follows the scaling behavior 
$P_{e,min} ( J_1, J_2, \sqrt{J_1 J_2})= P_{e,min} (1,J_2/J_1,\sqrt{J_2/J_1})$. 
More generally the probability of error is such that 
\begin{align}\label{sc1}
P_{e} (k J_1,k J_2, k h)= P_{e} (J_1,J_2,h) \qquad \forall k > 0 \:.
\end{align}
Upon exploiting this scaling and fixing $J_1=1$ we can study $P_e$ at 
$\tilde{h}$ as a function of $J_2\equiv J$. The behavior of $Q(J)\equiv 
P_{e,min}(1,J, \sqrt{J})$ is illustrated in the left panel of Fig. 
\ref{Pe234}. The  function has a cusp 
in $J=1$, whereas the tails of the curve for $J\rightarrow 0$ and
$J\rightarrow \infty$ go to zero faster with increasing size. 
This means that as the number of spins increases, the overlap between 
two different ground states approaches to zero. According to the scaling 
in Eq. (\ref{sc1}) the relevant parameter is the ratio between the two 
couplings and not the absolute difference. In turn, this means that $Q(J)$ 
is symmetric around $J=1$ in a log-linear plot. 
Expanding $Q(J)$ around $J=1$ and $J=0$ we obtain the following behavior
\begin{align}\label{sc2}
Q(J) & \stackrel{J\simeq 1}{=}\frac{1}{2} - \alpha_L \left| J-1 \right| 
+ O\left|J-1\right|^2 \\ 
Q(J) & \stackrel{J\rightarrow 0}{=}\frac{1}{2} - A_L + \beta_L \sqrt{J} 
+ \gamma_L J + O(J^{3/2}) \nonumber 
\end{align}
where $\alpha_L \in (0,1/2)$ is an increasing function of $L$.
According to the scaling (\ref{sc1}) the behavior of $Q(J)$ for
large $J$ is obtained by the replacement $J\rightarrow 1/J$ in the
second line of Eq. (\ref{sc2}). The parameters $A_L$, $\alpha_L$,
$\beta_L$, and $\gamma_L$ are reported in Table. \ref{t:pars} for
$L=2,3, 4$.  The corresponding Chernoff bound $\xi_J=-\log
\left[4\,Q(J)(1-Q(J))\right]$ does not carry additional information 
about the discrimination problem, but exhibits a simpler behavior
\begin{align}\label{sc2a}
\xi_J& \stackrel{J\simeq 1}{=} \frac{\delta_L}{16} \left| J-1 \right|^2 + O\left| J-1\right|^3 \\ 
\xi_J& \stackrel{J\rightarrow 0}{=} L \log 2 -L \sqrt{J} + \frac{L}{2}
J+ O(J^{3/2})\:, \nonumber
\end{align}
where $\delta_L = L!/4L$ for $L=3,4$ and half of this value for $L=2$.
The behavior of $\xi_J$ for large $J$ is again obtained 
by replacing $J\rightarrow 1/J$ in the second line of Eq. (\ref{sc2a}).
In the right panel of Fig. \ref{Pe234} we show 
$\xi_J$ as a function of $J$ for $L=2, 3, 4$.
\begin{table}[h]\label{t:pars}
\caption{Parameters $A_L$, $\alpha_L$, $\beta_L$, and $\gamma_L$ 
appearing in Eq. (\ref{sc2}), {\em i.e} the expansion of the rescaled
probability of error $Q(J)$ around $J=0$ and $J=1$.}
\begin{tabular}{l|l|l|l|l}
$L$ & $\alpha$ & $\beta$ & $\gamma$ & $A$ \\
\hline
2 & $\alpha_2=1/8=0.125$ &
$\beta_2 = 1/2\sqrt{2}\simeq 0.354$ &
$\gamma_2 = 1/4\sqrt{2}\simeq 0.177$ &
$A_2 = 1/2\sqrt{2}\simeq 0.354$
\\
3 & $\alpha_3=\sqrt{3}/8\simeq 0.217$ &
$\beta_3 = \sqrt{3}/8\simeq 0.217$ &
$\gamma_3 = 5 \sqrt{3}/32\simeq 0.271$ &
$A_3 = \sqrt{3}/4\simeq 0.433$
\\
4 & $\alpha_4\simeq 0.306$ &
$\beta_4 = 1/2\sqrt{14}\simeq 0.134$ &
$\gamma_4 = 23/28 \sqrt{14}\simeq 0.220$ &
$A_4 = \sqrt{14}/8\simeq 0.468$ \\
\hline 
\end{tabular}
\end{table}
\par
As mentioned in the previous Section, when we compare ground states of 
Hamiltonians with infinitesimally close values of the coupling $J$, the 
proper measure to be considered is the QCB metric, with the point of maximal
discriminability of two states corresponding maxima of the QCB metric tensor.  
At zero temperature $ds_{QCB}^2 =ds_B^2$ and thus \cite{JIsing}
\begin{eqnarray} 
g_J &=&\frac{h^2}{4 (h^2 + J^2)^2}, \quad L =2 \nonumber \\
g_J &=&\frac{3 h^2}{16 (h^2 - hJ + J^2)^2}, \quad L =3 \nonumber \\
g_J &=&\frac{h^2 (h^4 + 4h^2J^2 + J^4)}{4 (h^4+J^4)^2}, \quad L =4 
\label{gtz} \,.
\end{eqnarray}
Notice the simple scaling $g_J (k J, k h) = g_J (J,h)$, which is valid 
$\forall L$. Maxima of $g_J$ are thus obtained for $h^*=J$ for
$L=2,3,4$, and actually this is true for any $L$ (see also the next Section). 
The pseudo-critical point  $h^*$ which maximizes the QCB metric, turns 
out to be independent of $L$ and  equal to the true critical point, $h_c= J, 
\, \forall L$. At its maximum $g_J$ goes like $1/J^2$ which means that
it is easier to discriminate two infinitesimally close ground states for 
small $J$ rather than for large ones.
\subsection{Large $L$}
For large $L$, the overlap (fidelity $F$) between two different ground states
$|\psi_k\rangle\equiv |\psi_0 (J_k)\rangle$, $k=1,2$
is given by
\begin{align}\label{overlap} 
F=\braket{\psi_1}{\psi_2}= \prod_k
\cos \frac{\theta_{1k} -\theta_{2k}}{2} 
\end{align} 
where $k=(2n+1)\pi/L$ and $n$ runs from $1$ to $L/2$. Obviously, 
$F=1$ if $J_1=J_2$. Otherwise, one has $\cos[(\theta_{1k} -\theta_{2k})/2]<1$ 
and the fidelity $F$ quickly decays as the ratio of the couplings is 
different from one.  Solving $\partial_h \cos[(\theta_{1k} -\theta_{2k})/2]=0$ 
one finds that the overlap has a cusp in $\tilde{h}=\pm \sqrt{J_1 J_2}$, 
where it achieves the minimum value, corresponding to the minimum of the 
probability of error $P_e$.
In the thermodynamic limit $L\to\infty$, the overlap between two
different ground states goes to zero no matter how small is the
difference in the parameters $J_1$ and $J_2$. In other words, 
the different ground states become mutually orthogonal, a behavior 
known as orthogonality catastrophe \cite{pzPRE06}. In the critical 
region, corresponding to the vanishing of one of the single particle 
energies $\epsilon_{k}^2 + \Delta_{k}^2 =0$ with $k= 2\pi/L$, this 
behavior is enhanced, occurs for smaller $L$, and corresponds to a 
drop in the fidelity even for small values of $|J_2-J_1|$.
\par
For what concerns the QCB metric, upon taking the limit $T\to 0$ 
in Eq.(\ref{dsB}), we have that the classical part $ds_c^2$, which 
depends only on thermal fluctuations, vanishes due to the factor 
of $(\cosh (\beta\Lambda_k/2))^{-2}$. Therefore, at zero temperature, 
only the nonclassical part of Eq.(\ref{dsB}) survives and one obtains
$g_J =\frac{1}{4} \sum_{k } (\partial_J\vartheta_k)^2$,
where
$$\partial_J \vartheta_k =\frac{1}{1+ (\Delta_k /\epsilon_k )^2} 
(\partial_J \frac{\Delta_k}{\epsilon_k})= \frac{- h\sin
k}{\Lambda_k^2}\:.$$ 
Since we are in the ground state, the allowed quasi-momenta are 
$k = \frac{(2 n+1)\pi}{L}$ with $n=0,\ldots, L/2-1$.  Explicitly we have
\begin{align}\label{g_JTzero}
g_J =\frac{1}{4}\sum_k \frac{h^2\sin (k)^2}{\Lambda_k^4}.
\end{align} 
We are interested in the behavior of the QCB metric in the
quasi-critical region,  which is described by small values of the scaling 
variable $z \equiv L (h-J) \simeq L/\xi$, that is $z\approx 0$. 
Conversely the off-critical region is given by $z\rightarrow\infty$.
We substitute $h =  J+ z/L$  in Eq.(\ref{g_JTzero}) and expand around 
$z=0$ to obtain the scaling of $g_J$ in the quasi-critical regime
$g_J = \frac{1}{4}\sum_{k_n}\frac{(J+\frac{z}{L})^2\sin^2 (k_n)}{[
\frac{z^2}{L^2} + 4J(J+\frac{z}{L})\sin ^2(k_n/2)]^2} \equiv \sum_{k_n} f_{k_n}(z) $.
Since $\partial_z f(0)=0$,  the maximum of $g_J$ is always at $z=0$ for all 
values of $L$, in turn, the pseudo-critical point is $h^*= J = h_c$ $\forall L$.  
Going to second order and using Euler-Maclaurin formula, we get
\begin{equation}
g_J = \frac{L^2}{4} \left(\frac{1}{8 J^2}  - \frac{z^2}{384 J^4} \right)  
-\frac{L}{8 J^2} + O(L^0)\:,
\end{equation}
which shows explicitly that at $h=J$ the QCB metric has a maximum and 
there it behaves as
\begin{equation}\label{xx2}
g_J \simeq \frac{L^2}{32 J^2} + O(L)\:.
\end{equation}
From Eq. (\ref{xx2}) one concludes that the $1/J^2$ scaling of the
metric may be compensated by using long chains, which thus appears as
the natural setting to address the discrimination problem for large $J$.
\section{Discrimination of thermal states}\label{Tfinite}
In this section we address the problem of discriminating two states 
at finite temperature, {\em i.e.} we consider two thermal states
of the form $\rho_J = Z^{-1} e^{-\beta H(J)}$, $Z= \Tr[e^{-\beta
H(J)}]$, and analyze the behavior of the error probability, the 
Chernoff bound and the Chernoff metric as a function of the temperature 
and the external field. We discuss short chains $L=2,3,4$ and then 
the case of large $L$.
\subsection{Short Ising chains $L=2, 3, 4$}
\begin{figure}[h]
\includegraphics[width=0.45\textwidth]{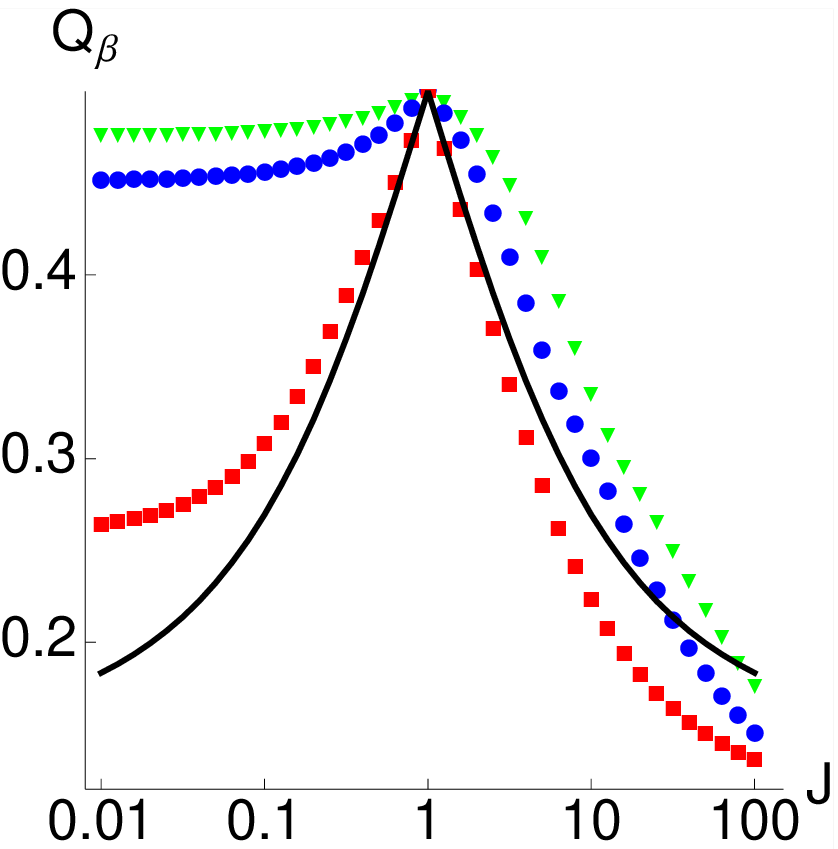}
\includegraphics[width=0.45\textwidth]{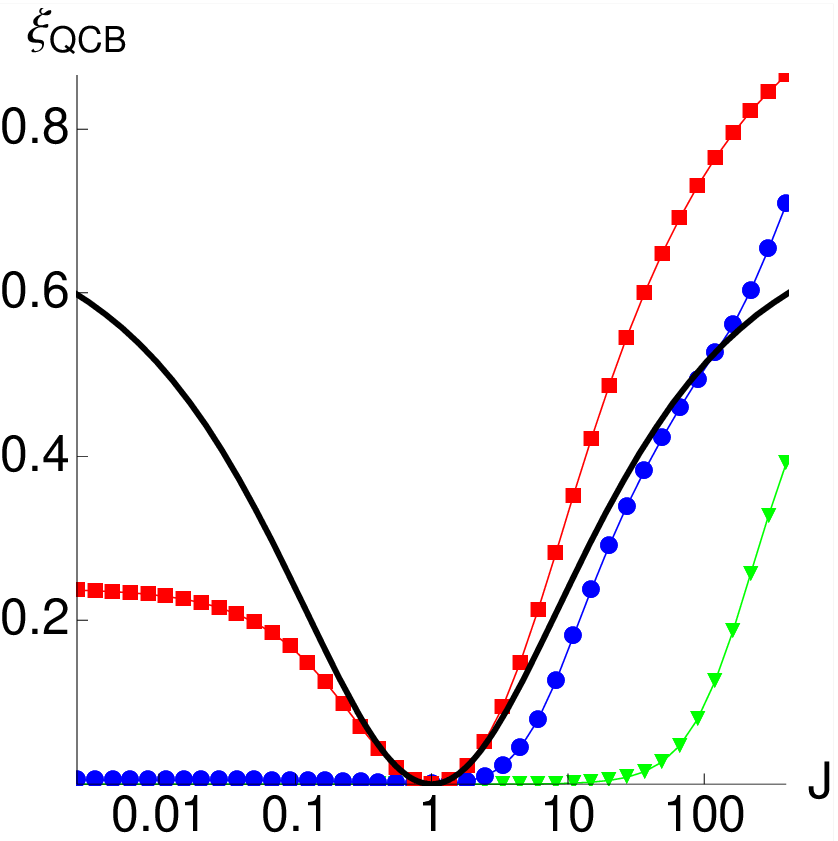}
\caption{(Left): Log-linear plot of the rescaled minimum probability of 
error $Q_\beta(J)\equiv P_{e,min}(1,J,\sqrt{J},\beta)$ for $L=2$ as a 
function of $J$. Green triangles correspond to $\beta=0.05$, blue
circles to $\beta=0.1$ and red squares to $\beta=1$. The black solid
curve is the probability of error in the zero temperature case.
The main effect of temperature is the loss of symmetry around 
$J=1$. (Right): Log-Linear plot of the quantum Chernoff bound $\xi_{QCB}$ for $L=2$. 
Green triangles correspond to $\beta=0.05$, blue circles to $\beta=0.1$ 
and red squares to $\beta=1$. We also report the zero temperature QCB 
for comparison (solid black curve).}\label{f:Pe2T}
\end{figure}

\begin{figure}[h]
\includegraphics[width=0.45\textwidth]{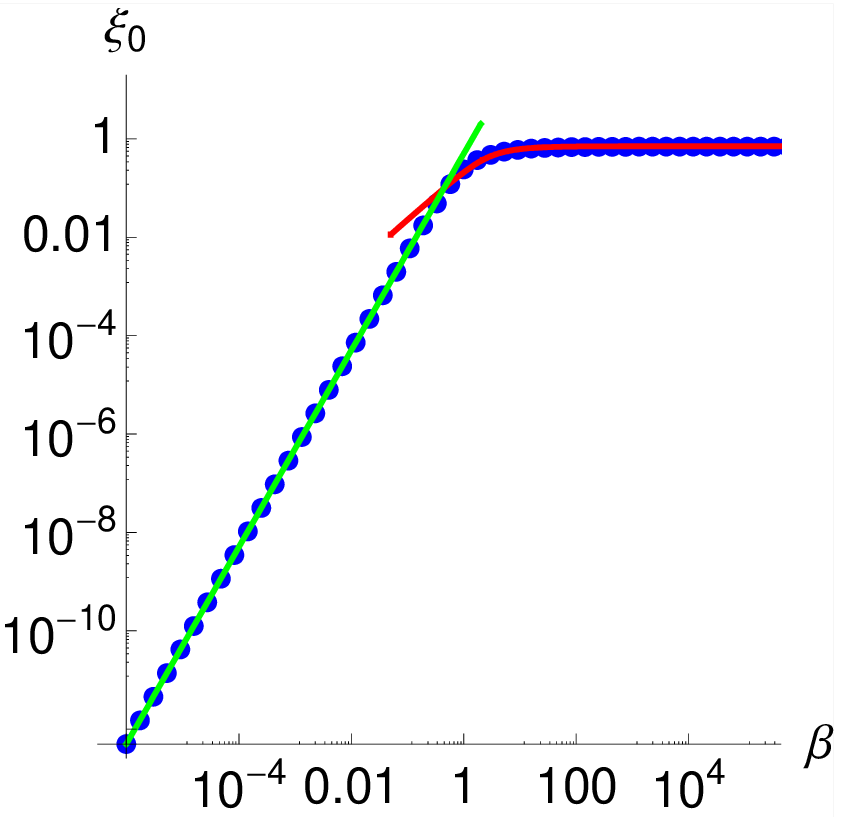}
\includegraphics[width=0.45\textwidth]{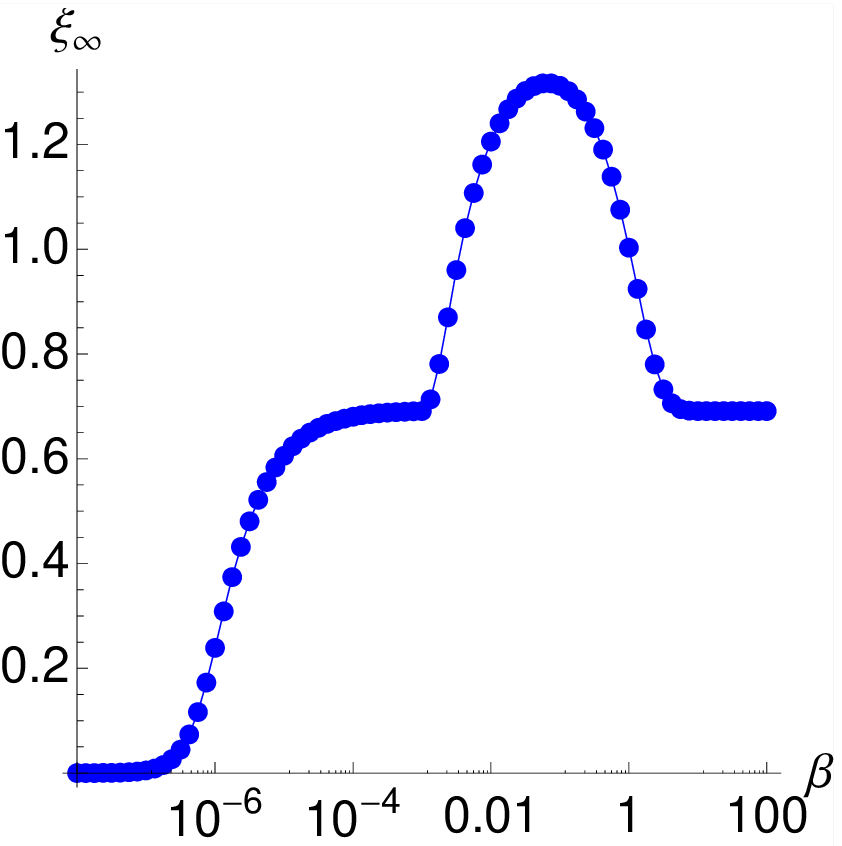}
\caption{(Left): Log-log plot of the Chernoff bound for vanishing $J$, 
$\xi_0 \equiv \xi_{QCB}(1,J\rightarrow 0, \sqrt{J},\beta)$,
as a function of inverse temperature $\beta$ (blue points)
together with the approximating functions of Eq. (\ref{fitxi01}) (green
line) and (\ref{fitxi02}) (red line).
(Right): Log-linear plot of the Chernoff bound for diverging $J$, 
$\xi_\infty \equiv \xi_{QCB}(1,J\rightarrow \infty, \sqrt{J},\beta)$,
as a function of inverse temperature $\beta$ 
}\label{f:asym}
\end{figure}

\begin{figure}[h]
\includegraphics[width=0.43\textwidth]{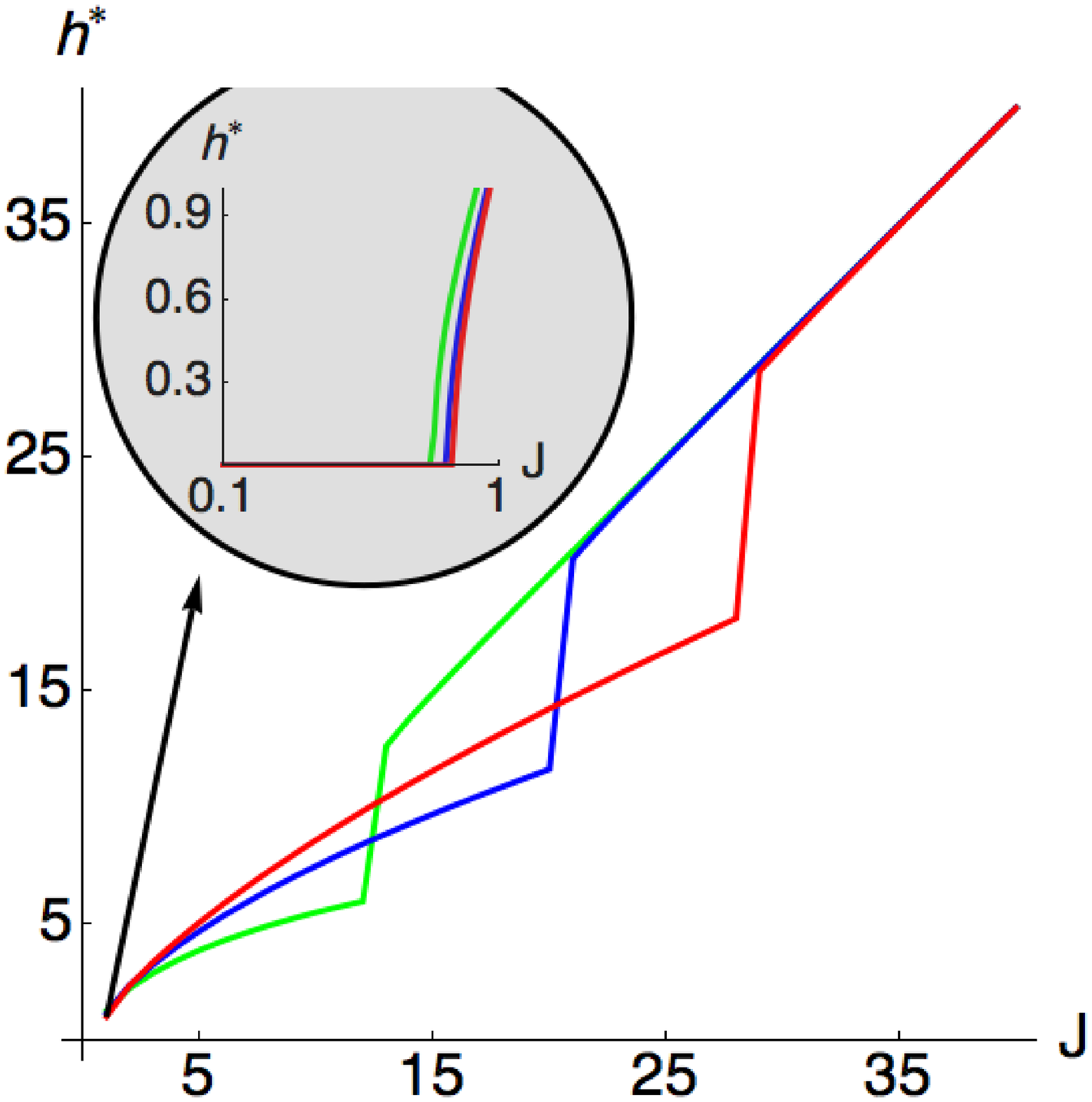}
\includegraphics[width=0.43\textwidth]{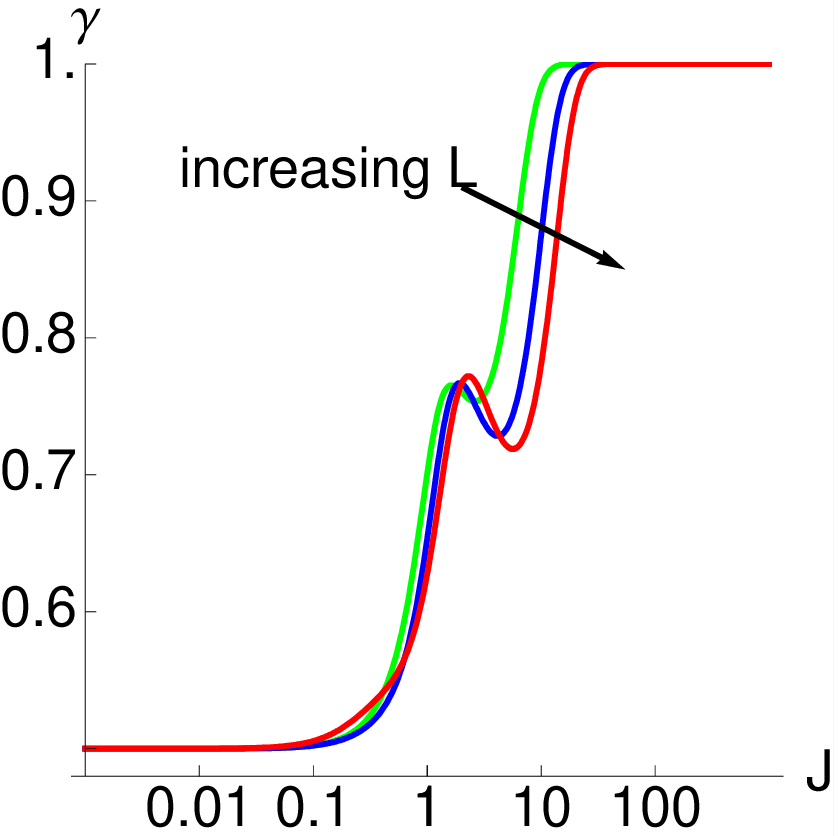}
\caption{(Left): linear plot of the optimal field $h^*$ maximizing the 
QCB  metric as  a function of $J$ for $\beta=1$. The inset shows the 
region of small $J$. (Right): log-linear plot of the ratio $\gamma$ 
between the (maximized) QCB and Bures metrics as a function of $J$ for 
$L=2,3,4$ (green, blue and red lines respectively) and $\beta=1$.}
\label{f:hscb}
\end{figure}

For short chains we have evaluated the probability of error by explicit
diagonalization of $\rho_{2}-\rho_{1}$, with $\rho_k\equiv\rho_{J_k}$. 
The probability of error 
follows the scaling
\begin{align}\label{sc3}
P_e(k J_1, k J_2,k h, \beta/k)= P_e (J_1,J_2,h,\beta)\:,
\end{align}
which may be exploited to analyze its behavior upon fixing $J_1=1$.
The main difference with the zero temperature case is that 
the error probability does depend on the absolute difference
between the two couplings, and not only on the ratio between them.
The optimal field $\tilde{h}$, minimizing $Q_\beta(J)=P_e(1,J,\tilde{h},\beta)$ 
is zero for small $J$, then we have a transient behavior and finally, 
for large $J$, $\tilde{h}=\sqrt{J}$. The range of $J$ for which 
$\tilde{h}\simeq 0$ increases with temperature (small $\beta$).  
In the left panel of Fig. \ref{f:Pe2T} we compare $Q_\beta(J)$ for 
$L=2$ and different values of $\beta$ to the analogous zero temperature 
quantity $Q_\infty (J)$. As it is apparent from the plot the main effect 
of temperature is the loss of symmetry around  $J=1$. Analogous behavior 
may be observed for larger $L$. Notice that discrimination at finite
temperature is not necessarily degraded.
\\
Upon diagonalization of the Hamiltonian we have also evaluated the quantum 
Chernoff bound by numerical minimization of 
$\min_s
\Tr\left[\rho_1 ^s\:\rho_2^{1-s}\right]$ and obtained for
$\xi_{QCB}$ the same scaling properties (\ref{sc3}) observed for the
error probability.
In the right panel of Fig. \ref{f:Pe2T} we compare the QCB for 
$L=2$ and different
values of $\beta$ to the analogous zero temperature quantity. 
Again the main effect of temperature is the loss of symmetry around  
$J=1$. Analogous behavior may be observed for larger $L$.
For vanishing $J$ the Chernoff bound $\xi_{QCB}(1,J\rightarrow 0, \sqrt{J},\beta)\equiv
\xi_0$ saturates to a limiting value scaling
with $\beta$ as 
\begin{align}
\label{fitxi01}
\xi_0 &\simeq \beta^2/2  
\quad\qquad\qquad\qquad\beta\rightarrow 0 \\
\xi_0 &\simeq \frac{\sqrt{2}}{\pi}\arctan
(\beta/2)  \qquad \beta \rightarrow \infty\:.
\label{fitxi02}
\end{align}
On the other hand, for diverging $J$ $\xi_{QCB}(1,J\rightarrow \infty, 
\sqrt{J},\beta)\equiv \xi_\infty$ shows the non monotone behaviour
illustrated in the right panel Fig. \ref{f:asym}. In the left panel
we report $\xi_0$ as a function of $\beta$ together with the
approximating functions of Eqs. (\ref{fitxi01}) and (\ref{fitxi02}). Overall, we notice that 
both for the single-copy and many-copy case, increasing the temperature 
may also results in an improvement of  discrimination, at least in the 
region of large couplings and intermediate temperatures.
\\
Finally, we have evaluated the QCB metric and found that it follows
the scaling
\begin{align}\label{sc4}
g_J(J,h,\beta)=\beta^2 \Phi_L(\beta J,\beta h)
\end{align}
where the form of the function $\Phi_L$ depends on the size only. 
The same scaling is also true for the Bures metric with different
functions $\Phi_L$. Indeed, this behavior follows directly from 
the common structure of the two metrics and by the fact that $g_J$ 
is obtained from the square of the derivative with respect to $J$.
The scaling is actually true for any size $L$. 
The optimal value $h^*$ of the external field, which maximizes the QCB 
metric at fixed $J$ and $\beta$ may be found numerically. Upon exploiting
the scaling properties we consider $\beta=1$ and found that $h^*$ is zero 
for small $J$, then we have a transient behavior and finally, 
for large $J$, $h^*=J$. According to the scaling above, the range of $J$ 
for which $h^*\simeq 0$ increases with temperature 
(small $\beta$) and viceversa.
In turn, for $\beta\rightarrow\infty$ we recover the results of the
previous Section, {\em i.e.} the critical point is always the optimal
one for discrimination.  
This behavior is illustrated in the left panel Fig. \ref{f:hscb}, where 
we report the optimal field $h^*$ as a function of $J$ for $\beta=1$. 
The inset shows the small $J$ region.
As we have noticed in the previous Section the two metrics are equal 
in the zero temperature limit. For finite temperature this is no longer
true and a question arises on whether the whole range of values allowed
by the inequality $\frac{ds_B^2}{2}\leq ds_{QCB}^2 \leq ds_B^2$
is actually spanned by the QCB metric. This is
indeed the case, as it may be seen by analyzing the behavior of the
ratio $\gamma=ds^2_{QCB}/d^2_{B}s$ at the (pseudo) critical point $h^*$
(we take the maximum of both the metrics, which generally occurs at
different values of the field).
In the right panel of Fig. \ref{f:hscb} we report $\gamma$ as a function 
of $J$ for $\beta=1$ and $L=2,3,4$. As it is apparent from the plot, for 
small $J$ we have $ds^2_{QCB} \simeq \frac12 ds^2_{B}$, whereas for large 
$J$ the two quantities become equal $ds^2_{QCB} \simeq ds^2_{B}$. The ratio 
is not monotone and the dependence on the size is weak. Upon exploiting the 
scaling in Eq. (\ref{sc4}) we may easily see that the range of $J$ for which 
the two metrics are almost equal increases with $\beta$. For vanishing 
temperature ($\beta\rightarrow\infty$) $ds^2_{QCB} \simeq ds^2_{B}$
everywhere and we recover the results of the previous Section. Conversely, 
for high temperature we have $ds^2_{QCB} \simeq \frac12 ds^2_{B}$ also for 
very large $J$. Also the transient region is shrinking for increasing
temperature.
\subsection{Large $L$}
In the limit of large size $L$ the behavior of the Chernoff metric follows 
the same scaling of Eq. (\ref{sc4}) found for short chains. The optimal value 
of the field which maximizes the QCB metric is $h^\ast=J$ for any finite 
temperature, where the metric element has a cusp. We have studied the 
QCB metric in the quantum-critical region $\beta |J-h| \ll 1$ and for 
low temperature $T\to 0$. The classical elements of the metric vanish due 
to the factor  $1/\cosh^2(\beta \Lambda_k/2)$ and we are left to analyze 
the nonclassical part $g_J^{nc}$ as a function of $T$. Bounding the metric 
by  functions that have the same scaling behavior in $\beta$ \cite{abasto}, 
will ensure that the metric itself scales with the same exponent.
The dispersion relation is linear around $k=0$ and we approximate 
$\Lambda_k \sim J k$ at the critical point $J=h$. Upon defining
$$
f(\beta,k)= \left \{ \begin{array}{ll} \beta^2 k^2 /4 & 0\leq k \leq 2/\beta \\
 1 & 2/\beta \leq k \leq \pi \end{array} \right.\:, $$
we have, for all $\beta$ and $k$, $\frac12 f(\beta,k)< \tanh^2(\beta Jk/2) 
< f(\beta,k)$. For large $L$, the sum on the classical part of the QCB metric 
may be replaced by the integral $ L\int dk$, thus leading to  
\begin{align}
g_J^{nc} \simeq \frac{L}{2\pi} \int_0^{2/\beta}dk 
\tanh^2{(\beta J k/2)}\frac{1}{J^2 k^2} 
+\frac{L}{2\pi} \int_{2/\beta}^\pi dk \tanh^2{(\beta 
 \Lambda_k/2)}\frac{J^2\sin^2(k)}{\Lambda_k^4}. \label{int2}
\end{align}
This is a good approximation in the limit $\beta\to\infty$ because the
upper integration limit $2/\beta$ becomes arbitrarily close to
$0$.  The first integral is bounded by $\frac{L}{2\pi}
\int_0^{2/\beta}dk \frac{f(\beta,k)}{2}\frac{1}{J^2 k^2}\leq
\frac{L}{2\pi} \int_0^{2/\beta}dk \tanh^2{(\beta J k/2)}\frac{1}{J^2
k^2} \leq \frac{L}{2\pi} \int_0^{2/\beta}dk f(\beta,k)\frac{1}{J^2
k^2}$.  The bounding integrals scale as $L\beta$ and the first integral
must scale in the same way for $\beta\to\infty$.  The
second term is upper bounded by $\frac{L}{2\pi}
\int_{2/\beta}^\pi dk \tanh^2{(\beta \Lambda_k/2)}\frac{J^2
\sin^2(k)}{\Lambda_k^4} \leq \frac{L}{2\pi} \int_{2/\beta}^\pi dk
\frac{1}{J^2 k^2} \sim L \beta $.  Therefore, since the bounding
integral scales as $\beta L$, $g_J^{nc}$ must scale as  $\beta L$ to the
highest order.  Observe that in the quantum-critical region $g_J \sim L$
is extensive, whereas at the critical point it has a superextensive
behavior $g_J \sim L^2$.  The nonclassical element scales algebraically
with temperature and in the zero temperature limit it diverges, matching
the ground state behavior that we described in the previous Section.
These results remark that criticality provide a resource for quantum
state discrimination, and that the discrimination of quantum states is
indeed improved upon approaching the QCP. 
\section{conclusions}\label{conclusions}
We have addressed the problem of discriminating two ground states or 
two thermal states corresponding to different values of the coupling 
constant in the one-dimensional quantum Ising model. We have
analyzed both short and long chains with the aim of assessing
the role of criticality (pseudo criticality for short chains) 
in single-copy and many-copy discrimination as well as in the
discrimination of infinitesimally closed states.
\par
At zero temperature both, the error probability for single-copy 
discrimination, and the Chernoff bound for $n$-copy discrimination 
in the asymptotic limit, are optimized by choosing the external field 
as the geometric mean of the two (pseudo) critical points. In this 
regime, the relevant parameter governing both quantities is the ratio 
between the two values of the coupling constant. On the other hand, 
the Chernoff metric is equal to the Bures metric and is maximized at 
the (pseudo) critical  point.
For finite temperature we have analyzed in some details the scaling 
properties of all the above quantities and have derived the optimal 
external field. We found that the effect of finite temperature is
twofold. On the one hand, critical values of the field are optimal
only for large values of the coupling constants. On the other hand, 
the ratio between the couplings is no longer the only relevant parameter
for both the error probability and the Chernoff bound, which also 
depends on the absolute difference. The ratio between the 
Chernoff metric and the Bures metric decreases continuously, but not
monotonically, for increasing temperature and approaches $1/2$ in the 
limit of high-temperature.
\par
In conclusion, upon considering the one-dimensional Ising model as
a paradigmatic example we have quantitatively shown how and to which 
extent criticality may represent a resource for state discrimination 
in many-body systems.
\section*{Acknowledgments}
The authors thank Paolo Giorda, Lorenzo Campos Venuti, Marco Genoni, Paolo 
Zanardi, and Michael Korbman for stimulating discussions.

\end{document}